\documentclass[conference]{IEEEtran}
\IEEEoverridecommandlockouts
\usepackage{cite}
\usepackage{amsmath,amssymb,amsfonts}
\usepackage{algorithmic}
\usepackage{graphicx}
\usepackage{textcomp}
\usepackage{xcolor}
\usepackage{booktabs}
\usepackage{multirow}
\usepackage{array}

\def\BibTeX{{\rm B\kern-.05em{\sc i\kern-.025em b}\kern-.08em
    T\kern-.1667em\lower.7ex\hbox{E}\kern-.125emX}}
\begin{document}

\title{A Survey of Bluetooth Indoor Localization\\
{\footnotesize \textsuperscript{}}
\thanks{}
}

\author{\IEEEauthorblockN{1\textsuperscript{st} Taolei Shi}
\IEEEauthorblockA{\textit{School of Computer Science and Technology} \\
\textit{University of Science and Technology of China}\\
Hefei, China \\
shitaolei99@mail.ustc.edu.cn}
\and
\IEEEauthorblockN{2\textsuperscript{nd} Wei Gong}
\IEEEauthorblockA{\textit{School of Computer Science and Technology} \\
\textit{University of Science and Technology of China}\\
Hefei, China \\
weigong@ustc.edu.cn}

}

\maketitle

\begin{abstract}
Nowadays, indoor localization has received extensive research interest due to more and more applications' needs for location information to provide a more precise and effective service \cite{7219371,7563896}. There are various wireless techniques and mechanisms that have been proposed; some of them have been studied in depth and come into use, such as Wi-Fi, RFID, and sensor networks. In comparison, the development of Bluetooth location technology is slow and there are not many papers and surveys in this field, although the performance and market value of Bluetooth are increasing steadily. In this paper, we aim to provide a detailed survey of various indoor localization systems with Bluetooth. In contrast with the existing surveys, we categorize the exciting localization techniques that have been proposed in the literature in order to sketch the development of Bluetooth location compared to other technologies. We also evaluate different systems from the perspective of availability, cost, scalability, and accuracy. We also discuss remaining problems and challenges to accurate Bluetooth localization. 
\end{abstract}


\section{Introduction}
\IEEEPARstart{I}{n} many cases, obtaining locational information about people or objects will facilitate our lives, such as looking for a lost phone, providing navigation services, and so on. Over the last two decades, positioning targets has become a very fascinating field of research with the expeditious maturation of IoT \cite{8874972,8844779,8752525}. Today, some theories of localization have been put into use, and some techniques have shown good performance. 

In the open air, GPS (Global Positioning System) has already got sufficient accuracy to meet the vast majority of user needs. But GPS cannot be applied in a room because the device cannot receive the signal. Additionally, the user's need for indoor localization varies from case to case. For example, a product recommendation system wants quicker positioning results in order to let users close to its shop receive the marketing information. It does not care too much about the accuracy compared with an indoor navigation system. And applications used for finding lost objects usually tend to achieve a higher coverage rate relative to error magnitude because the positioning result will be meaningless or even misleading if the lost item is not in the area indicated \cite{7420939}.

Indoor localization has various application demands and hence, at present, there is no wireless technique and mechanism like GPS performed indoors, which attracts researchers to explore further. Bluetooth, as one of the most widely used wireless communication technologies, is certainly taken into consideration.

Bluetooth (or IEEE 802.15.1) is designed for connecting different fixed or mobile wireless devices within a certain personal space. Bluetooth Low Energy (BLE), the latest version of Bluetooth, has shown good performance. Its maximum speed and coverage can be up to 24 Mbps and 100 meters, respectively, with low power consumption \cite{7120085}.

Today, Bluetooth indoor localization has obtained several results and been applied to different scenarios. In the next section, Section II, we illustrate some preliminary technical information regarding indoor positioning. In Section III, we discuss the problems and challenges Bluetooth indoor localization systems are dealing with. Next, in Section IV, we discuss the existing approaches for indoor positioning in the literature. Then in Section V, we compare the different positioning approaches and solutions. Then, we analyze the solutions in Section IV to prospect the development direction of positioning in the indoor environment in Section VI. Finally, in Section VII, we render our conclusion.

\section{Preliminaries}
Here we present some preliminary information that researchers must be aware of regarding indoor positioning systems and environments.
\subsection{Types of Localization}
There are many solutions which can be used in indoor localization. They can be divided into three types: triangulation, scene analysis, and proximity \cite{5483748}. These types are almost universal, not just for Bluetooth.

\begin{figure}[!t]
  \centering
  \includegraphics[width=2.5in]{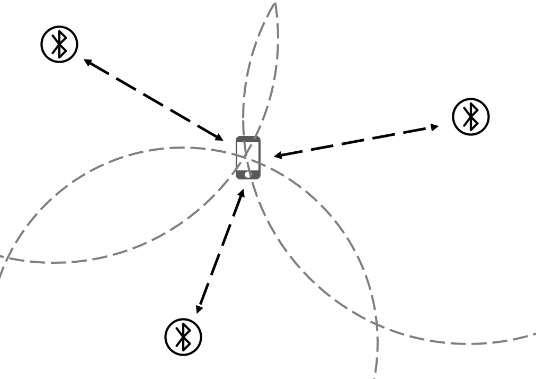}
  \caption{The type of triangulation.}
  \label{fig_1}
  \end{figure}

The triangulation converts the signal information to geometric values, such as the distance or angle \cite{7275525,9209674}. Then the principle gets the target position by geometric calculation. The triangulation needs at least three anchor nodes and the corresponding coordinates of each node in the location area, as shown in Fig.\ref{fig_1}.

The scene analysis uses fingerprinting to locate the target. A fingerprint is a feature that distinguishes different coordinates within a scene. A fingerprint consists of the environmental characteristics collected at the corresponding coordinates, which usually are collections of radio signals received from anchor points in the scene. Before the principle starts to locate the target, we need to collect scene information and build a database for feature matching \cite{5483748}.

The proximity is mainly used to estimate the target’s location based on the signal value. We use antennas that are fixed in the scene to cover the whole area. The target equipped with a corresponding device will be considered at the location of the antenna which receives the strongest signal during detection.

\subsection{Localization Techniques}
There are various signal metrics which are used for localization. In this section, we mainly introduce those already applied in Bluetooth indoor localization.

\subsubsection{RSSI}The received signal strength (RSS) based approach is widely used for indoor localization and the most of existing Bluetooth positioning solutions use RSSI (received signal strength indicator) value as the input for localization. The RSS can be used to estimate the distance between the transmitter and the receiver based on the signal propagation model \cite{1651798}. There are several signal propagation models for use, and parameters in each model can vary according to different environments. 

RSS-based localization can use three types of principles mentioned above. In the case of triangulation, the RSS can be used to calculate the distance between the target and the anchor point. So, as long as there are three or more reference nodes, we can use geometric calculation to obtain the location of the target with tag \cite{9209674}. In the case of scene analysis, the RSSI at each anchor point is applied to a certain location to obtain the feature of scene analysis. Different positions should detect different RSSI values at some reference points in view of the relationship between RSSI value and distance \cite{ke2018developing}. In the case of proximity, we need a single anchor point to create a geofence as a basis for judgement. And then obtain the proximity of the user to the anchor by estimating the distance with a path loss function.

While the RSSI-based approach is widely used in Bluetooth localization, it shows poor position accuracy in general. There are many obstacles in the complex indoor environment and they will influence the transmission of signals, which makes signal propagation models calculate the distance inaccurately. In addition, the RSS is not stable and usually fluctuates widely because of multipath effects and signal noise.

\subsubsection{CSI}In many wireless systems, such as IEEE802.11 and UWB, the coherence bandwidth of the wireless channel is smaller than the bandwidth of the signal, which makes the channel frequency selective and different frequencies present different amplitude and phase behavior. In addition, in multiple antennae pairs, the channel frequency responses may significantly vary with the antennae distance and signal wavelength, which is different from the RSS. The RSS merely provides an estimate of the average amplitude over the whole signal bandwidth and the accumulated signal over all antennae, which makes the RSS susceptible to multipath effects and interference and causes severe fluctuation over time. By contrast, the channel state information can obtain both the amplitude and phase responses of the channel at different frequencies and between separate transmitter-receiver antennae pairs.

The CSI-based approach is rarely used in Bluetooth localization because of the limitations of the PHY protocol. BLE uses Gaussian Frequency Shift Key (GFSK) modulation for transmitting data. In traditional frequency shift keying (FSK) protocols, each symbol is represented as a frequency \cite{ayyalasomayajula2018bloc}. So when transmitting different information, the frequency will vary over time. And due to the Gaussian filter, the variation of frequency becomes continuous, which means phase measurement becomes an issue. As a result, CSI can not be easily captured on BLE.

\subsubsection{Fingerprinting}Scene analysis based positioning techniques need environmental preprocessing before the system is ready to be used \cite{9023649}. Environmental information should be collected to obtain fingerprints or features during an offline phase in order to build a database of fingerprints. After the system is deployed, the measurements obtained during the online phase are compared with fingerprints in the database to estimate the location of the tag. RSSI and CSI are usually captured for obtaining fingerprints or features. There are many algorithms that can be used to match the online measurements with fingerprints in a database, which can be probabilistic methods or approaches that involve machine learning.

A fingerprinting-based approach is common in Bluetooth localization. The Euclidean distance can be used as the basis of fingerprint matching to develop a beacon-based location system for smart home device management. Multiple neural networks are used as the matching algorithm to construct a positioning resolution, allowing indoor navigation with a cost-effective Bluetooth architecture.

\subsubsection{Other methods}In addition to the localization techniques mentioned above, BLE can also be used with Angle of Arrival (AoA) and Time of Flight (ToF), which is relatively rare in Bluetooth localization.

AoA-based approaches utilize antenna arrays to estimate the angle of the signal from the transmitter to the receiver. When the transmitted signal arrives at the antennae array at the receiver, there is a time difference between the individual elements of the antennae array capturing the signal, which is relative to the angle of arrival. So AoA takes advantage of this feature to obtain the estimated angle by calculating the time difference. AoA can estimate the location of a tag with as few as two reference nodes and is susceptible to multipath effects.

ToF utilizes the signal propagation time to calculate the distance between the transmitter and the receiver. We can think of the distance as the result of the ToF value multiplied by the speed of light. The ToF needs at least three reference nodes to locate the target device. After the calculated distance between each reference node and the tag, the location of the target can be estimated through a geometric formula. ToF requires strict synchronization between nodes and the device with a tag, which the BLE protocol does not include.

\section{Problems and challenges}
In this section, we discuss the existing problems and challenges that need to be dealt with when constructing Bluetooth indoor localization systems and algorithms.
\subsection{Accuracy}
The accuracy of positioning is one of the most important characteristics that influences the performance of the localization system. The indoor environment is a challenging space to locate the target. Compared to the outdoors, there are obstacles and multipath effects which affect the transmission and reception of signals negatively. Therefore, it is a problem that needs to be overcome to limit the impact coming from indoor environments, such as multipath effects \cite{ayyalasomayajula2018bloc} and other noises, when improving the accuracy of positioning systems. Now, there are some approaches that can be used to mitigate the environmental influence. LocBLE \cite{chen2017locating} passes raw RSS data through an adaptive noise filter(ANF) which contains two noise filtering techniques. Zhou \cite{9209674} and Bluepass \cite{5546506} use a machine learing algorithm to process RSS data.

The design of BLE also has a certain impact on localization accuracy. In order to achieve the longest possible battery life, the BLE protocol design specifies a series of power-saving features that make the BLE protocol ultralight, which leads to the large fluctuations of RSS in dynamic propagation environments \cite{chen2017locating,faragher2014analysis}.

\subsection{Latency}
BLE has a decent data rate of 24Mbps, which can achieve fast data transmission within the coverage range of 100m \cite{7120085}. So, theoretically, with suitable methods, a Bluetooth localization system can be designed to achieve near real-time positioning, even in real-time. Since the accuracy of Bluetooth positioning is still a great need to improve, researchers often choose to sacrifice latency for accuracy if necessary when exploiting the positioning approaches, such as extra movement or stillness of the target for the sake of the localization algorithm. But after approaches achieve a certain precision, meeting the needs of the positioning system, the latency will be an inescapable problem when deploying the system.

The challenge of latency asks for localization systems to be capable of locating the target with as little delay as possible, so as not to be noticed by users. This means the system should be able to operate the algorithm to implement localization with a limited amount of captured signal information within a short time. Therefore, there is a need for optimized signal processing at an acceptable cost of time.

\subsection{Range of Coverage}
The coverage area that the localization system can be up to is also very important in actual use. The increased reception range means fewer reference nodes are required in larger indoor environments when deploying the positioning system, which is appealing to large indoor spaces such as hospitals, malls, and jails. But as the distance between the transmitter and receiver increases, the quality of the signal decreases. This is not only affected by the hardware itself, but also by the interference of the indoor environment. It is obvious that a localization system will perform worse in more complex environments \cite{ke2018developing}. The coverage range is constrained by accuracy in this case.

\subsection{Cost}
The cost of positioning systems consists of the equipment cost and the set-up cost. The ideal system should not use any additional equipment that is high-end and not widely used, which is not only too expensive but also not conducive to popularity, although the use of some special equipment can improve the accuracy to a certain extent. Additionally, the set-up cost can not be neglected either. For example, building the database of fingerprints or obtaining the parameters of the signal propagation model needs the collected information about the environment before the system comes into use, which takes time and manpower. And these behaviors may need to be repeated to maintain the accuracy of localization when obstacles in the indoor environment change. 

\subsection{Security}
The security of Bluetooth localization systems has not been attractive in the current research. But Bluetooth localization systems certainly need to deal with the situation that users in the personal network locate the target in private space, which makes it an inevitable problem in business.

\section{Literature review}
Researchers have made substantial contributions to the topic of Bluetooth indoor localization systems. There are many meaningful contributions in the Bluetooth positioning area that deal with different application scenarios. In this section, we discuss the state of the art of positioning systems with different approaches. These approaches show different adaptability to environmental changes in practical applications, which can be divided into three categories: totally unaffected methods; weakly affected methods; and strongly affected methods.

\subsection{Totally unaffected methods}
Some approaches use geometric knowledge and novel algorithms to achieve locations whose accuracy is basically unaffected by the environment. These systems do not need to be trained for each new environment and retrained when the environment changes.

Bluepass \cite{5546506} and BLoc \cite{ayyalasomayajula2018bloc} do not need environmental information in the process of localization. It means that they can maintain the positioning accuracy for a long time without maintenance. And it is one of the necessary steps to enable zero startup cost Bluetooth localization.

Bluepass focuses on the construction of the overall structure of the positioning system. It details the operation of the central server and the local server in charge of the corresponding room in the scene and provides an architecture of interactions between the user and target among different rooms. Limited by the shortcomings of the algorithm and the unreliability of RSSI, Bluepass shows low accuracy. CSI is not available directly from BLE, which causes the sparse use of Bluetooth localization, although CSI contains more signal information than RSS. Ayyalasomayajula proposed BLoc that overcomes the issue of CSI acquisition and achieves a localization accuracy of 86 cm in a multipath-rich environment.

Bluepass is an indoor localization method based on a voting scheme that eliminates cells close to an anchor point, detecting the target with a weak signal. Bluepass proposed a complete structure dealing with positioning objects in indoor spaces with several rooms. First, Bluepass elects the room where the user is probably located by the Single Coverage Density Method (SCDM). SCDM divides the entire scene into a grid with its cells initialized with zero. According to the RSSI, each anchor creates a square and increases the value of cells within this square. Thus, the cell contained in more squares has a bigger number, which means a higher signal density. The intersection of the squares will finally provide the area with the highest signal concentration. The room, including this area, is voted to the next step. Bluepass chooses the ITU model to calculate the distance based on RSSI in order to fit the environment. Then Bluepass obtains the specific location of the target by the triangulation principle, which achieves an average error of 3.23m in the whole area. 

BLoc mainly tackles three problems that prevent CSI based localization on BLE devices: access to CSI, narrow bandwidth, and multipath resolution. According to the PHY protocol, BLE uses Gaussian frequency shift keying (GFSK) modulation for transmitting data. Each symbol is represented as a frequency in the traditional frequency shift keying (FSK) protocols. Due to the Gaussian filter, the change in frequency becomes continuous, which makes it hard to measure the signal phase because this procedure needs a steady frequency. LocBLE leverages a simple technique that is to transmit BLE data packets with constant long sequences. In this way, channels with a constant frequency for a period of time can be obtained and the signal phase can be measured.

BLE’s effective bandwidth is narrow, so it is difficult for BLE to separate out different paths. The multipath effect has a negative impact on localization accuracy, so it is an inevitable problem when improving positioning accuracy. LocBLE utilizes frequency hopping to combine frequency bands for a wider effective bandwidth. Then the system can have the basic ability to deal with multipath effects. LocBLE tends to choose the shortest path and computes the spatial entropy of the likelihood distribution around all peaks that correspond to the direct and indirect paths. These two factors are combined to determine the direct path to reduce the impact of multipath effects.

\subsection{Weakly affected methods}
Some methods collect environmental information to train algorithms or assist localization before systems are deployed in a new environment. It does not need to be done again for each environmental change. Such systems are usually updated periodically to maintain accuracy because the impact of environmental change on the system is not explicit enough.

\begin{figure}[!t]
  \centering
  \includegraphics[width=2.5in]{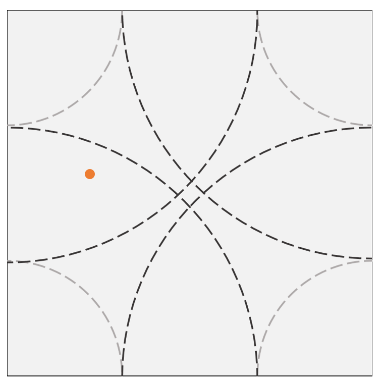}
  \caption{Generation of small regions.}
  \label{fig_2}
  \end{figure}

Wang \cite{7420939} proposed two BLE-based localization schemes, Low-precision Indoor Localization (LIL) and High-precision Indoor Localization (HIL), utilizing the collected RSSI measurements to generate a small region with the goal of including the target as much as possible, which is different from most of the positioning methods that attempt to obtain a more accurate single location of the target, as shown in Fig.\ref{fig_2}. It is applicable to some scenarios, such as finding lost objects in private space, that are concerned more with the effectiveness of positioning than accuracy. Zhou \cite{9209674} proposed a self-adaptive Bluetooth indoor localization system using an LSTM-based distance estimator. Zhou summed up the challenges propagation models are dealing with in two points: the fast-changing radio environment and the issues of hardware devices. In order to reduce the impact, the system uses a new distance estimator based on deep learning models. Mustafa \cite{mustafa2021high} proposed a positioning system for multi-story buildings that used part of the iBeacon frame fields, which is difficult to imitate by other techniques. In addition, the system needs a map to assist in localization.

The success rates of HIL can exceed 90 percent, and LIL can be up to 100 percent, although the system is only suitable for a small number of applications \cite{7420939}. Fluctuations in RSSI have an impact on the division of radius and the final results. The different areas, which have different sizes and derive from different classification rules in the system, also make the high success rate a little less stable. Zhou \cite{9209674} used an LSTM-based distance estimator instead of a propagation model. In fact, the distance estimator plays the role of a propagation model in the system. The distance estimator needs to be well trained before deployment in a new environment. To some extent, self-adaptive mechanisms adopted by the system mitigate the impact of environmental changes. The navigation system in \cite{mustafa2021high} optimized the positioning accuracy of users in motion by about 2 meters. It utilizes the major and minor fields that are each 16 bits long to store location information of nodes in multi-story buildings and a "snap to path" technique to improve the location accuracy.

The fundamental basis of LIL and HIL in \cite{7420939} is the relationship between the RSS value and distance, which has been obtained by measurements before. They draw circles for each anchor point, and the radius of each circle depends on the RSS values. A simple distribution of anchor points with circles can divide the scene into several regions. LIL only determines the size of the radius by whether the nodes receive RSSI signals or not at two levels of transmission power. Each anchor point has two circles, and the smaller one corresponds to the level of lower power. 

On the basis of LIL, HIL attempts to find a potentially predictive relationship between RSSI value and distance. HIL collects a bunch of RSSI samples every meter and then does a cluster analysis in order to find the most suitable partition sample with RSSI and corresponding distance. HIL finds the partition radiuses in both high and low power levels and then divides more regions in the scene into smaller areas, which means a lower successful rate covering the target than LIL. Obviously, a single environmental change has a certain impact on the partition radius of HIL, but it cannot directly influence the final result.

A long short-term memory (LSTM) network is used to address the time dependency of sequential measured RSSI data in \cite{9209674}. The latest element of LSTM output and other features are further processed by a fully-connected multilayer perceptron neural network and a non-linear sigmoid activation function. After the distance estimator is well trained, the system optimizes the triangulation by a series of self-adaptive mechanisms. Elastic radius intersecting is used to make sure the circles of adjacent anchor nodes intersect in pairs, which can influence the next step of multiple weighted-centroid localization. In the actual positioning process, the circles of nodes are usually not exactly tangent. In acceptable form, the location of the target is contained in the triangle whose vertexes are achieved by the intersection of the circles generated by three adjacent nodes. Elastic radius intersecting is the guarantee that makes this situation stable. 

Then the system in \cite{9209674} selects the top K nodes whose measured RSSIs are the strongest among all nodes. Three of them can be combined to obtain one point of weight-centroid localization. These estimations of position can be averaged with the location of the target at the last moment to gain a reference point which is resistant to high variance of measurements and can be used to exclude the more distant nodes among the K nodes in order to further rule out the possibility of abnormal RSS values. Then the rest nodes can be used to obtain the point of multiple weighted-centroid localization. Self-adaptive Kalman Tracking is used to predict the movement of a target in order to address unpredictable fluctuations in measured RSSI during a short period of time. The system optimizes the prediction of propagation and triangulation and, as a result, achieves 1.5 m localization accuracy in warehouse scenarios.

The size of two iBeacon frame fields is enough for thousands of beacons to store the exact location. These fixed anchor nodes can be divided into detailed categories, which makes anchor points in the space quite explicit for localization \cite{mustafa2021high}. The "snap to path" technique makes sure the target cannot be located in any inaccessible places in space. The target displayed on the map is always on the path it travels. Since the map is coded into the system, what needs to be done is to switch the coordinates of the estimated location, which is outside the limit of the map, with the closest location on the path the target moves. This method mitigates the effect of random fluctuations in RSS values.

The system in in \cite{mustafa2021high} selected five different smoothing algorithms for data collection. The initial raw RSSI values are filtered through a Kalman filter in order to smooth the noisy nature of BLE signals. After that, the data should be filtered by one of the smoothing algorithms, which depends on the filtering effect. Then the data is sorted through a weighted function to determine the proximity of each beacon. Through a path-loss model, the location of the target can be estimated by triangulation based on the distance corresponding to the RSS value since the locations of fixed nodes are already captured. While the user is in motion, it keeps a window that records the beacons closest to the target, which makes the algorithm able to determine the direction of travel by calculating the enhancement and attenuation of the signal of anchor nodes.

The above methods do not pay attention to environmental changes in the process of positioning but need to rely on specific environmental information in the deployment stage. In general, environmental information plays an auxiliary role in the system and can even be directly diverted if the scenes are too similar.

\subsection{Strongly affected methods}
Some approaches tend to collect all the necessary information in the indoor environment, such as fingerprints. They must update the database regularly to maintain the accuracy of positioning. The accuracy of environmental information greatly affects the accuracy of localization.

Altini \cite{5483748} explored the matching algorithm for fingerprinting. Altini proposed a localization system with multiple neural networks that handles the changes in RSSI values due to user orientation and designed a predictive connection system that can speed up the localization process. Brouwer \cite{de2021optimal} optimized the receiver location for passive indoor positioning. The system is based on an inverse fingerprinting model. In contrast to the common fingerprinting technique where the target device receives signals from fixed reference points, the inverse model lets the target device broadcast the signal to each anchor node, and the RSS values contained within a fingerprint are all from the target device during the same period. The set of collected fingerprints forms the radio map for the match. LocBLE \cite{chen2017locating} applies fingerprints to its own algorithm, which makes it different from fingerprinting. LocBLE is a special localization method that is different from the common patterns. Most approaches utilize the fixed anchor nodes to locate the target with a tag, such as a mobile phone. But LocBLE in turn locates the BLE node in the scene with the smart phone users carrying it with them.

The two systems in \cite{5483748,de2021optimal} both show poor performance, but for different reasons. Because the main goal is to realize indoor navigation, Altini placed only a few anchor points in one room, which makes the positioning accuracy inadequate. Brouwer paid attention to the location of nodes and achieved a better scheme for node placement, but he ignored the main problems fingerprinting dealt with.

In \cite{chen2017locating}, the user's device plays the role of the observer in the process of localization, and the target is the stationary or moving BLE node in the scene. LocBLE determines the status of the target and detects the observer’s (and the target’s if moving) step and direction. LocBLE requires users to complete an L-shaped movement for locating BLE beacons and handling symmetry ambiguity. Then LocBLE adopts the environment classification to judge whether to start a new regression or not. If the environment remains unchanged, LocBLE chooses to continue the regression by appending the new data. The regression originates from the combination of the path-loss model and the distance equation between the BLE beacon and observer, and then reformulates them into the form of an elliptical regression problem.

LocBLE provides a new approach for searching and tracing with Bluetooth positioning systems and achieves an average of 1.8m accuracy in locating indoor BLE beacons.LocBLE requires users to move before locating the target and uses a unique algorithm, which cannot be divided into basic types, to achieve high accuracy. But it needs to collect fingerprints to monitor the environment in the algorithm, which means its accuracy is still sensitive to environmental changes.

\section{Discussion}
In this section, we compare different Bluetooth positioning systems. The comparison is mainly about how localization systems respond to challenges and problems, which are discussed above. 

Table \ref{tab:table1} shows a comparison among positioning systems that are picked from above. Note that in the table we refer to triangulation as (T) and scene analysis as (S). We notice that the accuracy of the system in \cite{7420939} is not available because it is designed for a high success rate. The result of positioning is a small area where the target is estimated. Compared to other systems, BLoc has higher accuracy because it is based on CSI instead of RSSI, which contains more channel information. Moreover, BLoc also needs additional preprocessing for obtaining CSI on BLE. The system in \cite{mustafa2021high} is designed to locate the target in the whole building. It utilizes fields in beacon data to mark each anchor node in order to get a rough idea of where the target is in the building. Bluepass uses the SCDM to distinguish the room with the target among all the rooms in the whole area. The design of cross-room positioning makes Mustafa’s system \cite{mustafa2021high} and Bluepass show lower accuracy compared to others because they ignore problems Bluetooth localization deals with in one room.

Machine learning has a wide range of applications in the field of indoor localization, both in terms of the processing of RSSI values and research on localization algorithms, which have yielded results, but not much in the field of Bluetooth localization. It is worth mentioning that there are still some results of neural network based Bluetooth localization \cite{borenovic2008utilizing, genco2005three, 10.4108/ICST.MOBILWARE2008.2901, 5483748, 7329709}. Some details of neural network based Bluetooth localization are given in \cite{genco2005three}. In fact as \cite{10.4108/ICST.MOBILWARE2008.2901} states, neural networks are not widely used in positioning and localization systems although the noisy environment that characterizes such systems, especially indoor, is well suited for fingerprinting systems based on neural networks. In \cite{borenovic2008utilizing}, the author developed a WLAN localization system. Altini proposed \cite{5483748} to solve the problem given by how RSSI values change depending on user orientation. Tuncer He proposed an Artificial Neural Network(ANN) model equipped with temperature sensors to detect the mobile phone location.

In complex environments, localization systems need to face more unpredictable challenges. The system in \cite{9209674} achieves around 0.9 m and 1.5 m localization accuracy in LAB and warehouse scenarios, respectively. Zhou's view is that wireless signals are reflected, diffracted or even absorbed by dense metal bars of storage racks and also blocked by huge amounts of stacked goods. In densely populated areas, the movement of obstacles is unpredictable, which greatly increases the difficulty of localization. In addition, the RSSI values are significantly influenced by user orientation \cite{5483748}. LocBLE exhibits relatively low accuracy in locating moving targets. First, the BLE signal blockage changes too fast for the environment classifier to capture during the human's walking. Second, the error in movement estimation accumulates, especially in estimating the moving direction of two users. This means it difficult to accurately describe the relative positions of two close targets. In multi-story buildings, unlike single-story interiors, complex buildings with multiple floors are more likely to achieve unreasonable positioning results, such as the air outside the escalators in shopping malls, which poses a great challenge to the practicality of the positioning method. The system in \cite{mustafa2021high} uses a "snap to path" technique to calibrate the positioning results. This means that this method requires an accurate building construction drawing to achieve the positioning function. In order to design a location system that can provide good accuracy while the user is in motion through multiple rooms, the placement of BLE beacons is worth discussing.

\begin{table}[!t]
\centering
\caption{Comparison of existing systems\label{tab:table1}}
\begin{tabular}{m{1cm}|m{1cm}|m{0.5cm}|m{1cm}|m{1.5cm}|m{1.5cm}}
\hline
\multirow{2}{*}{\textbf{System}} & \multicolumn{5}{c}{\textbf{Evaluation}}\\
\cline{2-6}
& \textbf{Accuracy} & \textbf{Type} & \textbf{Range} & \textbf{Robustness} & \textbf{Extra}\\
\hline
system in \cite{7420939} & N/A & N/A & room & high success rate & needs RSS measurement in advance\\
\hline
Bluepass & 2-3m & T & rooms in the whole area & the client-server architecture in rooms  & finds the room with target before positioning\\
\hline
System in \cite{9209674} & 1.5m & T & room & good performance  & builds Propagation model with LSTM\\
\hline
LocBLE & 1-2m & N/A & room & user’s movement for positioning  & locates the BLE node in the scene\\
\hline
BLoc & 0.86m & N/A & room & good performance  & uses CSI instead of RSSI\\
\hline
System in \cite{de2021optimal} & 3m & S & room & higher accuracy with better location of receivers  & generates inverse fingerprints\\
\hline
System in \cite{mustafa2021high} & 2-4m & T & building & location snapped to the map  & distinguishes Beacons by data\\
\hline
\end{tabular}
\end{table}

\section{Analysis of research area}
In this section, we discuss the possible improvements that can be made in the future. In our paper, we analyze Bluetooth indoor localization as a research area and the classification of indoor localization.

Throughout our research, we found many contributions in RSS based approaches. RSS based localization systems with BLE have been studied for almost 20 years. The method to mitigate the severe fluctuation of RSS values is common, but the propagation models are still taken seriously. It is difficult to accurately model indoor radio attenuation because of the dynamic and complex environments. RSS-based approaches may need to seek other points of penetration, such as handling multipath effects or optimizing the principle. The average accuracy of fingerprinting is lower than that of RSS. The matching algorithms are one of the issues that must be focused on, which has several resolutions. Fingerprinting needs pretreatment of the environment and is susceptible to the dynamic scene, which makes fingerprinting uncompetitive.

CSI-based approaches give a new way of thinking about Bluetooth localization. Compared with RSS-based approaches, we may reach the conclusion that the multipath effect is the challenge that cannot be bypassed in the pursuit of accuracy. Looking for Bluetooth-specific location methods may be able to add many enhancements.

\section{Conclusion}
In this paper, we reviewed the types of localization, Bluetooth localization techniques, and corresponding systems. We discussed the challenges and problems that BLE positioning deals with. We described the solutions of existing localization systems to these problems and speculate on what the future of location techniques might be. We believe that location techniques other than RSS and multipath effects are the main challenges localization should overcome with BLE.

\bibliographystyle{IEEEtran}
\bibliography{refe.bib}

\begin{thebibliography}{10}
\providecommand{\url}[1]{#1}
\csname url@samestyle\endcsname
\providecommand{\newblock}{\relax}
\providecommand{\bibinfo}[2]{#2}
\providecommand{\BIBentrySTDinterwordspacing}{\spaceskip=0pt\relax}
\providecommand{\BIBentryALTinterwordstretchfactor}{4}
\providecommand{\BIBentryALTinterwordspacing}{\spaceskip=\fontdimen2\font plus
\BIBentryALTinterwordstretchfactor\fontdimen3\font minus
  \fontdimen4\font\relax}
\providecommand{\BIBforeignlanguage}[2]{{%
\expandafter\ifx\csname l@#1\endcsname\relax
\typeout{** WARNING: IEEEtran.bst: No hyphenation pattern has been}%
\typeout{** loaded for the language `#1'. Using the pattern for}%
\typeout{** the default language instead.}%
\else
\language=\csname l@#1\endcsname
\fi
#2}}
\providecommand{\BIBdecl}{\relax}
\BIBdecl

\bibitem{7219371}
Y.~Zhang, M.~Qiu, C.-W. Tsai, M.~M. Hassan, and A.~Alamri, ``Health-cps:
  Healthcare cyber-physical system assisted by cloud and big data,'' \emph{IEEE
  Systems Journal}, vol.~11, no.~1, pp. 88--95, 2017.

\bibitem{7563896}
M.~Qiu, W.~Dai, and A.~V. Vasilakos, ``Loop parallelism maximization for
  multimedia data processing in mobile vehicular clouds,'' \emph{IEEE
  Transactions on Cloud Computing}, vol.~7, no.~1, pp. 250--258, 2019.

\bibitem{8874972}
K.~Gai, Y.~Wu, L.~Zhu, Z.~Zhang, and M.~Qiu, ``Differential privacy-based
  blockchain for industrial internet-of-things,'' \emph{IEEE Transactions on
  Industrial Informatics}, vol.~16, no.~6, pp. 4156--4165, 2020.

\bibitem{8844779}
K.~Gai, L.~Zhu, M.~Qiu, K.~Xu, and K.-K.~R. Choo, ``Multi-access filtering for
  privacy-preserving fog computing,'' \emph{IEEE Transactions on Cloud
  Computing}, vol.~10, no.~1, pp. 539--552, 2022.

\bibitem{8752525}
K.~Gai, K.~Xu, Z.~Lu, M.~Qiu, and L.~Zhu, ``Fusion of cognitive wireless
  networks and edge computing,'' \emph{IEEE Wireless Communications}, vol.~26,
  no.~3, pp. 69--75, 2019.

\bibitem{7420939}
Y.~Wang, Q.~Ye, J.~Cheng, and L.~Wang, ``Rssi-based bluetooth indoor
  localization,'' in \emph{2015 11th International Conference on Mobile Ad-hoc
  and Sensor Networks (MSN)}, 2015, pp. 165--171.

\bibitem{7120085}
F.~Zafari, I.~Papapanagiotou, and K.~Christidis, ``Microlocation for
  internet-of-things-equipped smart buildings,'' \emph{IEEE Internet of Things
  Journal}, vol.~3, no.~1, pp. 96--112, 2016.

\bibitem{5483748}
M.~Altini, D.~Brunelli, E.~Farella, and L.~Benini, ``Bluetooth indoor
  localization with multiple neural networks,'' in \emph{IEEE 5th International
  Symposium on Wireless Pervasive Computing 2010}, 2010, pp. 295--300.

\bibitem{7275525}
Z.~Jianyong, L.~Haiyong, C.~Zili, and L.~Zhaohui, ``Rssi based bluetooth low
  energy indoor positioning,'' in \emph{2014 International Conference on Indoor
  Positioning and Indoor Navigation (IPIN)}, 2014, pp. 526--533.

\bibitem{9209674}
Z.~Li, J.~Cao, X.~Liu, J.~Zhang, H.~Hu, and D.~Yao, ``A self-adaptive bluetooth
  indoor localization system using lstm-based distance estimator,'' in
  \emph{2020 29th International Conference on Computer Communications and
  Networks (ICCCN)}, 2020, pp. 1--9.

\bibitem{1651798}
K.~Wendlandt, M.~Berhig, and P.~Robertson, ``Indoor localization with
  probability density functions based on bluetooth,'' in \emph{2005 IEEE 16th
  International Symposium on Personal, Indoor and Mobile Radio Communications},
  vol.~3, 2005, pp. 2040--2044 Vol. 3.

\bibitem{ke2018developing}
C.-K. Ke, W.-C. Ho, and K.-C. Lu, ``Developing a beacon-based location system
  using bluetooth low energy location fingerprinting for smart home device
  management,'' in \emph{International wireless internet conference}.\hskip 1em
  plus 0.5em minus 0.4em\relax Springer, 2018, pp. 235--244.

\bibitem{ayyalasomayajula2018bloc}
R.~Ayyalasomayajula, D.~Vasisht, and D.~Bharadia, ``Bloc: Csi-based accurate
  localization for ble tags,'' in \emph{Proceedings of the 14th International
  Conference on emerging Networking EXperiments and Technologies}, 2018, pp.
  126--138.

\bibitem{9023649}
B.~Wang, H.~Zhu, M.~Xu, Z.~Wang, and X.~Song, ``Analysis and improvement for
  fingerprinting-based localization algorithm based on neural network,'' in
  \emph{2019 15th International Conference on Computational Intelligence and
  Security (CIS)}, 2019, pp. 82--86.

\bibitem{chen2017locating}
D.~Chen, K.~G. Shin, Y.~Jiang, and K.-H. Kim, ``Locating and tracking ble
  beacons with smartphones,'' in \emph{Proceedings of the 13th International
  Conference on emerging Networking EXperiments and Technologies}, 2017, pp.
  263--275.

\bibitem{5546506}
J.~J.~M. Diaz, R.~d.~A. Maués, R.~B. Soares, E.~F. Nakamura, and C.~M.~S.
  Figueiredo, ``Bluepass: An indoor bluetooth-based localization system for
  mobile applications,'' in \emph{The IEEE symposium on Computers and
  Communications}, 2010, pp. 778--783.

\bibitem{faragher2014analysis}
R.~Faragher and R.~Harle, ``An analysis of the accuracy of bluetooth low energy
  for indoor positioning applications,'' in \emph{Proceedings of the 27th
  International Technical Meeting of The Satellite Division of the Institute of
  Navigation (ION GNSS+ 2014)}, 2014, pp. 201--210.

\bibitem{mustafa2021high}
A.~Mustafa and E.~R. Sykes, ``A high fidelity indoor navigation system for
  users in motion using ble with beacons,'' 2021.

\bibitem{de2021optimal}
R.~de~Brouwer, J.~Torres-Sospedra, S.~T. Oliver, and R.~Berkvens, ``Optimal
  receivers location for passive indoor positioning based on ble,'' 2021.

\bibitem{borenovic2008utilizing}
M.~Borenovic, A.~Neskovic, D.~Budimir, and L.~Zezelj, ``Utilizing artificial
  neural networks for wlan positioning,'' in \emph{2008 IEEE 19th International
  Symposium on Personal, Indoor and Mobile Radio Communications}.\hskip 1em
  plus 0.5em minus 0.4em\relax IEEE, 2008, pp. 1--5.

\bibitem{genco2005three}
A.~Genco, ``Three step bluetooth positioning,'' in \emph{International
  Symposium on Location-and Context-Awareness}.\hskip 1em plus 0.5em minus
  0.4em\relax Springer, 2005, pp. 52--62.

\bibitem{10.4108/ICST.MOBILWARE2008.2901}
A.~Shareef, Y.~Zhu, and M.~Musavi, ``Localization using neural networks in
  wireless sensor networks.''\hskip 1em plus 0.5em minus 0.4em\relax ICST, 5
  2010.

\bibitem{7329709}
S.~Tuncer and T.~Tuncer, ``Indoor localization with bluetooth technology using
  artificial neural networks,'' in \emph{2015 IEEE 19th International
  Conference on Intelligent Engineering Systems (INES)}, 2015, pp. 213--217.

\end{thebibliography}

\end{document}